\documentclass[sn-mathphys-num]{sn-jnl}

\usepackage{graphicx}%
\usepackage{amsmath,amssymb,amsfonts}%
\usepackage{amsthm}%
\usepackage{mathrsfs}%
\usepackage[title]{appendix}%
\usepackage{xcolor}%

\begin{document}

\title{Investigating the excitation function of HBT radii for Lévy-stable sources}

\author*{\fnm{Máté} \sur{Csanád}}\email{csanad@elte.hu}
\author{\fnm{Dániel} \sur{Kincses}}\email{kincses@ttk.elte.hu}
\affil{\orgdiv{Department of Atomic Physics}, \orgname{ELTE Eötvös Loránd University}, \orgaddress{\street{Pázmány Péter sétány 1/A}, \city{Budapest}, \postcode{1117}, \country{Hungary}}}

\abstract{Contemporary heavy-ion physics research aims to explore the phase diagram of strongly interacting matter and search for signs of the possible critical endpoint on the QCD phase diagram. Femtoscopy is among the important tools used for this endeavor; there have been indications that combinations of femtoscopic radii parameters (referred to as HBT radii for identical boson pairs) can be related to the system's emission duration. An apparent non-monotonic behavior in their excitation function thus might signal the location of the critical point. In this paper, we show that conclusions drawn from the results obtained with a Gaussian approximation for the pion source shape might be altered if one utilizes a more general Lévy-stable source description. We find that the characteristic size of the pion source function is strongly connected to the shape of the source and its possible power-law behavior. Taking this into account properly changes the observed behavior of the excitation function.}


\maketitle

\section{Introduction}
\label{introduction}
Exploring the phase diagram of strongly interacting matter is a cornerstone of heavy-ion physics research~\cite{Achenbach:2023pba}. The quest to pinpoint the critical endpoint on the Quantum Chromodynamics (QCD) phase diagram is ongoing, and femtoscopy serves as a pivotal tool in this endeavor~\cite{Bzdak:2019pkr}. It has been demonstrated that femtoscopic radii parameters (also called HBT radii, after R. Hanbury-Brown and R. Q. Twiss~\cite{HanburyBrown:1952na}) extracted from two-particle interferometry measurements are sensitive to the emission duration of heavy-ion collisions~\cite{Chapman:1994yv}. In the RHIC Beam Energy Scan program~\cite{Bzdak:2019pkr}, a non-monotonic collision energy (and consequently baryochemical potential) dependence of these parameters was observed~\cite{STAR:2014shf}, which was suggested to be a sign of possible critical behavior~\cite{Lacey:2014wqa}. These measurements assumed a Gaussian distribution for the shape of the two-particle emission distribution. Recent results from both experiment~\cite{PHENIX:2017ino,Kincses:2024sin,CMS:2023xyd,NA61SHINE:2023qzr,Porfy:2024kbk} and phenomenology~\cite{Korodi:2022ohn,Kincses:2022eqq,Ayala:2023sbb,Csanad:2024hva} showed that the shape of the pion pair-source follows a Lévy-stable distribution~\cite{Csorgo:2003uv} and exhibits a power-law tail. While new measurements in this direction are ongoing, in this paper, we describe a method to estimate the effect of the different source shapes on the previous results, based on already published data. In particular, we demonstrate that using a Lévy-stable assumption for the pion pair source can significantly change the values and trends of the radii parameters as a function of collision energy, and thus the previous conclusions drawn from the Gaussian measurements might need to be reevaluated. It has to be noted that the interpretations of the non-monotonicities mentioned above were derived based on the Gaussian interpretation, thus additional phenomenological analyses, similarly to Ref.~\cite{Csorgo:2003uv}, are needed to interpret the collision energy dependence of radii of Lévy-stable sources.

\subsection{Basic definitions}
Utilizing the smoothness approximation~\cite{Pratt:1997pw}, where the members of the investigated particle pair have approximately identical momenta, the two-particle correlation function (as a function of their relative momentum $\boldsymbol{q}$ and average momentum $\boldsymbol{K}$) is defined as 
\begin{align}
    C_2(\boldsymbol{q},\boldsymbol{K}) = \int d^3rD(\boldsymbol{r},\boldsymbol{K})|\psi^{(2)}_{\boldsymbol{q}}(\boldsymbol{r})|^2,
\end{align}
where $D(\boldsymbol{r},\boldsymbol{K})$ is the pair source distribution (also called spatial correlation function or pair emission function), and $\psi^{(2)}_{\boldsymbol{q}}(\boldsymbol{r})$ is the symmetrized pair wave function~\cite{Lisa:2005dd}. In absence of final-state interactions, the modulus square of the wave function is simply ${1+\cos{(\boldsymbol{q}\boldsymbol{r}})}$. This implies that the correlation function can be expressed with the the Fourier-transform of the spatial correlation function via this simple relation:
\begin{align}
    C_2^{(0)}(\boldsymbol{q},\boldsymbol{K}) = 1+\widetilde{D}(\boldsymbol{q},\boldsymbol{K}), 
\end{align}
where the $(0)$ superscript denotes the lack of final-state interactions, and $\widetilde{D}$ denotes the Fourier transform of $D$ in its first variable.

If one assumes a shape or functional form for the pair source distribution, one can also calculate the shape of the correlation function. The dependence on $\boldsymbol{K}$ is then usually understood to be realized through the parameters of the source, such as its width. Subsequently, one can test the initial assumption via fits to experimentally measured momentum correlation functions. Recent experimental investigations~\cite{PHENIX:2017ino,CMS:2023xyd,NA61SHINE:2023qzr,Kincses:2024sin} showed that assuming a so-called Lévy-stable distribution (a generalization of the Gaussian distribution) for the shape of the source might provide a statistically superior description to the measured two-pion correlation functions, in the sense that a $\chi^2$-test provides an orders of magnitude higher p-value for fits based on Lévy-stable distributions (as shown in the appendix).

The three-dimensional, elliptically contoured Lévy-stable distribution is defined as~\cite{Nolan:2014}
\begin{align}\label{e:Levy3D}
    \mathcal{L}(\alpha,\boldsymbol{R^2};\boldsymbol{x})&=\frac{1}{2\pi^3}{\int}d^3\boldsymbol{\omega}\exp{\left(i\boldsymbol{\omega}\boldsymbol{x}{-}\frac{1}{2}|\boldsymbol{\omega}^T\boldsymbol{R^2}\boldsymbol{\omega}|^{\alpha/2}\right)},\\
     \textnormal{where }\boldsymbol{R^2} &= \begin{pmatrix}
R_o^2 & R_{os}^2 & R_{ol}^2\\
R_{os}^2 & R_s^2 & R_{sl}^2\\
R_{ol}^2 & R_{sl}^2 & R_l^2
\end{pmatrix}.
\end{align}
Here, $\boldsymbol{x}$ is the variable of the distribution, $\boldsymbol{R^2}$ is a positive definite matrix containing the HBT radii parameters (also called Lévy-scale parameters), $\alpha$ is the Lévy exponent, and $\boldsymbol{\omega}$ is an integration variable vector. In the above definition of $\boldsymbol{R^2}$ we already utilized the usual notation of the Bertsch-Pratt $out-side-long$ coordinate system~\cite{Bertsch:1988db,Pratt:1990zq}, where the $out$ direction represents the direction of the average transverse momentum of the particle pair, the $long$ direction is the beam direction, and the $side$ is perpendicular to both. In the case of an azimuthally integrated experimental analysis, it is usual to neglect the off-diagonal terms, and in the following we will also do this.

Utilizing such a distribution as the pair-source function, in the absence of final-state interactions, the correlation function takes the following form:
\begin{align}\label{e:Cq3D}
    C_2(q_o,q_s,q_l) = 1+\lambda \exp{\left(-|R_o^2q_o^2+R_s^2q_s^2+R_l^2q_l^2|^{\alpha/2}\right)}.
\end{align}

Here, we also introduced the $\lambda$ correlation strength or intercept parameter~\cite{Csorgo:1994in,Bolz:1992hc}, corresponding to the correlation value at zero relative momentum (in an interaction-free case). In the case of $\alpha = 2$, the shape of the pair-source and the correlation function is both a Gaussian. In the case of $\alpha < 2$, the pair-source exhibits a power-law behavior, and the correlation function takes the form of a stretched-exponential function.
Of course, in the case of an experimental analysis, one needs to properly account for the final-state Coulomb interaction. For Lévy-stable sources, details of such calculations were discussed in Refs.~\cite{Kincses:2019rug,Csanad:2019lkp,Nagy:2023zbg}; in the following, we use the results presented in Ref.~\cite{Nagy:2023zbg}.

\section{Scaling method}

In the first published experimental heavy-ion analysis utilizing a Lévy-stable description for the pion pair source~\cite{PHENIX:2017ino}, a scaling variable was found, defined as 
\begin{align}
    \widehat{R} = \frac{R}{\lambda\cdot(1+\alpha)},
\end{align}
where $R$ is the one-dimensional Lévy scale parameter. In Ref.~\cite{PHENIX:2017ino} the physical interpretation of the parameter is not given. Still, one can deduce that $R/\lambda$ is related to the area ``under'' the measured correlation function (between the data points and the unity line), while the $\alpha$-dependent part encodes the shape change. This expression is thus of mathematical nature: it captures the most important part of the data points, and connects the best fitting curves for various $\alpha$ values.

Thus when fitting the same correlation function with different fixed $\alpha$ values, this parameter turns out to be remarkably stable. This implies that if we know the $\lambda$ and $R$ results for a given $\alpha$ assumption, we can deduce the value of these for another $\alpha$ assumption. In particular, if the Gaussian parameters $R_{\rm G}$ and $\lambda_{\rm G}$ (obtained with an assumption of $\alpha=2$) are known and the $\lambda_{{\rm free} \alpha}$ and $\alpha$ values are also known for a free-$\alpha$ fit, then the value of $R_{{\rm free} \alpha}$ can be estimated based on
\begin{align}
    \widehat{R}_{\rm G} \equiv \frac{R_{\rm G}}{\lambda_{\rm G}\cdot(1+2)} =
    \widehat{R}_{{\rm free}\;\alpha} \equiv \frac{R_{{\rm free}\;\alpha}}{\lambda_{{\rm free}\;\alpha}\cdot(1+\alpha)}
\end{align}
as
\begin{align}
    R_{{\rm free}\;\alpha} =R_{\rm G}\frac{\lambda_{{\rm free}\;\alpha}\cdot(1+\alpha)}{\lambda_{\rm G}\cdot(1+2)}.
\end{align}

We shall note that while the $\lambda$ parameter has a theoretical interpretation based on the core-halo model and chaoticity (as discussed, e.g., in Ref~\cite{PHENIX:2017ino}), experimentally, it may depend on additional settings, such as particle identification. This, however, does not affect the procedure outlined here since the ``$\widehat{R}$-scaling'' estimation is performed within the conditions of one experimental apparatus.

One may ask, furthermore, why all the different fixed-$\alpha$ fits can be used when many of these are invalidated by the low confidence level stemming from the large $\chi^2$ values of these fits. However, while the radii and correlation strengths stemming from these fits do not characterize the data physically, the mathematical scaling relation still holds for the largest part of the possible $\alpha$ values.

\begin{figure}
	\centering 
	\includegraphics[width=0.70\textwidth]{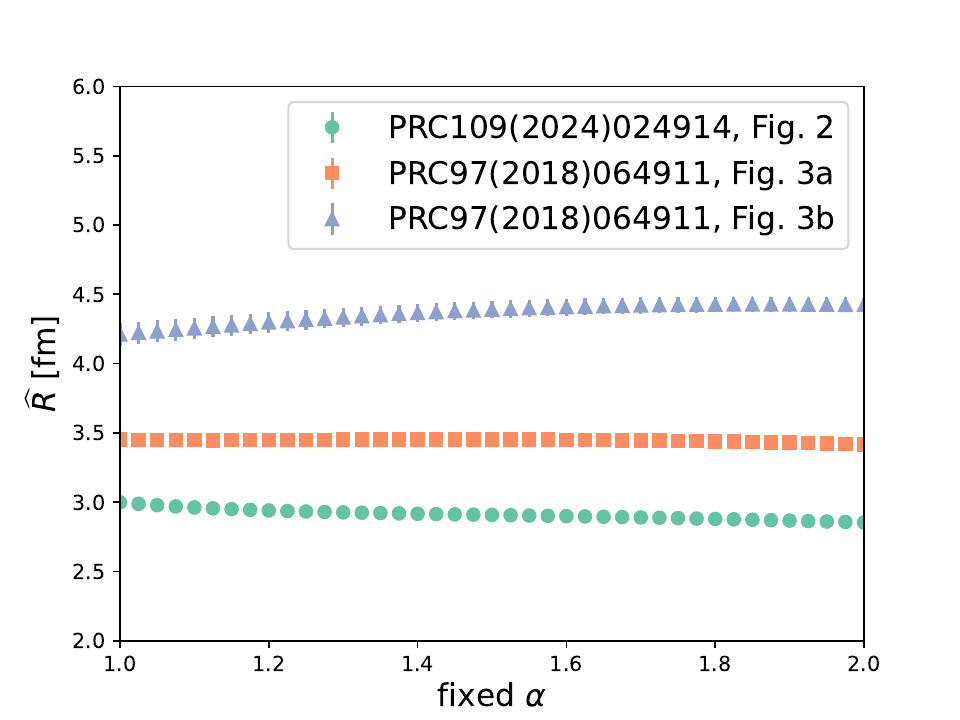}	
	\caption{Figure 2 of Ref. \cite{CMS:2023xyd} and Figs. 3a-3b of Ref. \cite{PHENIX:2017ino}. See detailed parameters in Appendix.} 
	\label{fig_rhat}%
\end{figure}

To demonstrate this and the stability of the $\widehat{R}$ scaling variable, we take three example correlation functions from Refs.~\cite{PHENIX:2017ino, CMS:2023xyd} (from their respective HEPdata entries). Using the correlation function calculation method described in Ref.~\cite{Nagy:2023zbg} and the software package given in Ref.~\cite{CoulCorrLevyIntegral}, we fit these example correlation functions with different fixed $\alpha$ values, with the functional form
\begin{align}
N  \left[ 1-\lambda  + \lambda  (1+e^{-(qR)^{\alpha}}) K_{C}(q;R,\alpha) \right](1+\epsilon q),
\end{align}
where $q$ is the one-dimensional momentum difference of the pair (usually taken in the longitudinally comoving frame or the pair rest frame), $K_C$ is the Coulomb correction as calculated in Ref.~\cite{Nagy:2023zbg}, $N$ is a normalization parameter and $\epsilon$ responsible for accounting for any residual, long-range background from non-femtoscopic effects (such as energy and momentum conservation, flow, or minijets). From the resulting fit parameters, we calculate the corresponding $\widehat{R}_{{\rm fix}\;\alpha}$ values. All fit parameters and fit quality indicators are shown in the Appendix. We find that, indeed, in the investigated range of $\alpha$ values, the scaling variable $\widehat{R}$ remains approximately constant as a function of the fixed $\alpha$ value, as demonstrated in Figure~\ref{fig_rhat}. A variation of up to 4-5\% is present, which may be interpreted as a systematic uncertainty of our analysis.

While an experimental confirmation is not yet available, assuming that this scaling relation also holds for the three-dimensional radii parameters is natural. This is because of its connection to the mathematical form of the fitted correlation functions: a smaller $\alpha$ leads to a more ``peaked'' correlation function; hence, to describe the same data, one requires a larger $R$ and a larger $\lambda$ value---this holds in three dimensions as well. Furthermore, as demonstrated in Ref.~\cite{Kurgyis:2018zck}, we assume that the $\alpha$ parameter measured under the assumption of a spherically symmetric source is identical with (or close to) the $\alpha$ obtained from a three-dimensional measurement (which does not assume a spherically symmetric source). Thus we can estimate how the radii would change from the fixed $\alpha=2$ Gaussian case to the free-$\alpha$ case using published data, as described in the next section.

\section{Results and discussion}

Measurements based on a Gaussian source assumption were performed in Ref.~\cite{STAR:2014shf}, where also the collision energy dependence of the source parameters (radii and $\lambda$) was investigated at $K_T\approx 0.22$ GeV$/c^2$, where $K_T$ is the transverse component of $\boldsymbol{K}$. In this paper, a non-monotonic collision energy dependence of $\sqrt{R_{\rm o,G}^2-R_{\rm s,G}^2}$ was found, as mentioned above.

\begin{figure}[t]
	\centering 
	\includegraphics[width=0.70\textwidth]{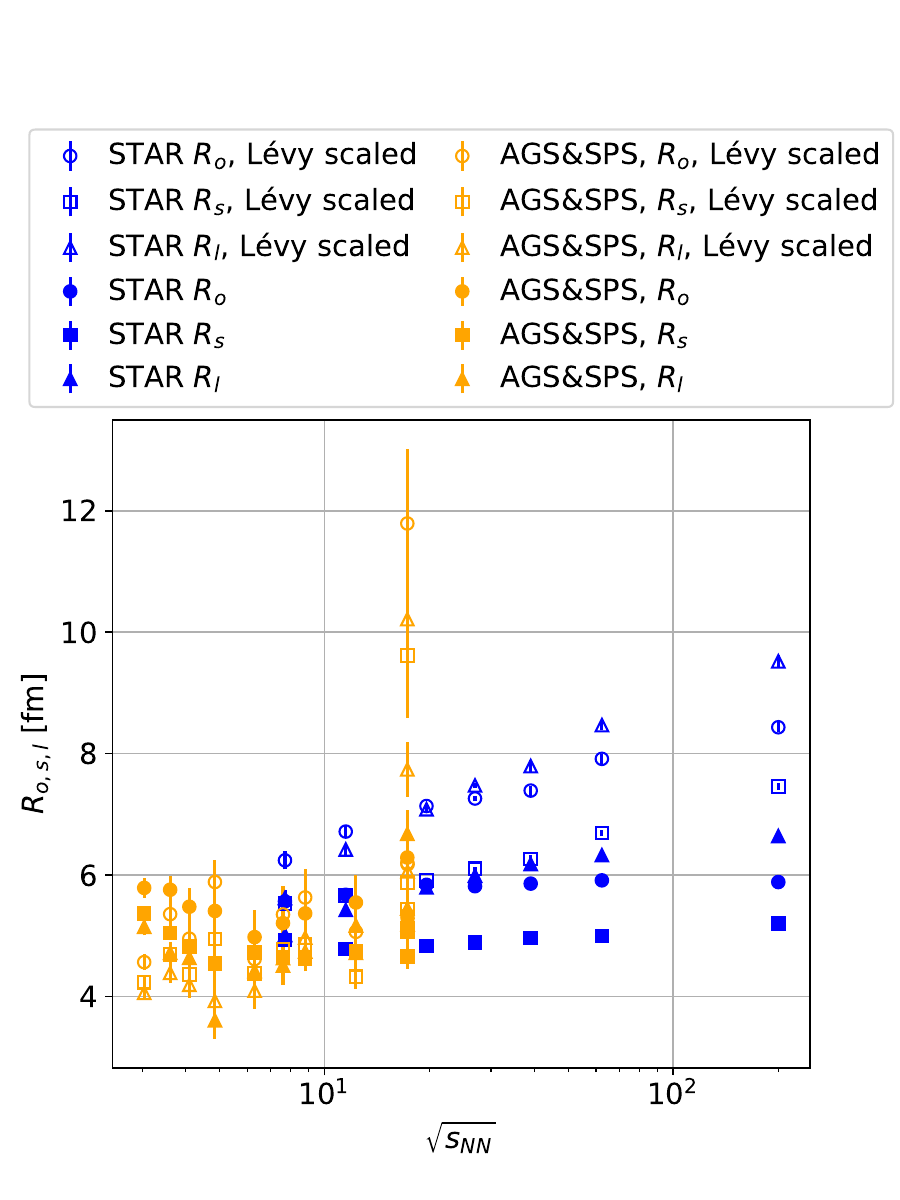}	
	\caption{Scaled Lévy parameters $R_{\rm o,s,l}$, compared to the original Gaussian values from Ref.~\cite{STAR:2014shf}} 
	\label{f:Ro2Rs2Levy}%
\end{figure}

Based on STAR preliminary results presented at the Zimányi School 2023 and CPOD 2024 conferences\footnote{See talk slides from the Zimányi Winter School 2023 at
\href{https://indico.cern.ch/event/1352455/contributions/5696657/}{indico.cern.ch/event/1352455/contributions/5696657/}
and from the CPOD 2024 at
\href{https://conferences.lbl.gov/event/1376/contributions/8829/}{conferences.lbl.gov/event/1376/contributions/8829/}}, yielding $\alpha$ values in the range of 1.3-1.7, one can make a rough approximation of the collision energy dependence of the $\lambda_{{\rm free}\;\alpha}$ and $\alpha$ parameters (also at $K_T\approx 0.22$ GeV$/c^2$) in the form of
\begin{align}
\alpha(\sqrt{s_{NN}}) & = A_{\alpha}\left(\frac{\sqrt{s_{NN}}}{E_0}\right)^{B_{\alpha}},\\
\lambda(\sqrt{s_{NN}}) & = A_{\lambda}\left(\frac{\sqrt{s_{NN}}}{E_0}\right)^{B_{\lambda}},    
\end{align}
with
\begin{align}
A_\alpha = 1.85, B_\alpha = -0.06,\;\;
A_\lambda = 0.6, B_\lambda = 0.06,
\end{align}
and $E_0 = 1$ GeV is introduced to remove the unit of $\sqrt{s_{NN}}$. Note that these relations are based on 1D measurements and may not be valid in the case of 3D measurements. It has, however, been observed~\cite{Kurgyis:2018zck} that 1D and 3D measurements yield within uncertainties the same $\alpha$ and $\lambda$ parameters, justifying the usage of the above relations here as well.

The Gaussian measurements for $R_{G(o,s,l)}$ together with the corresponding $\lambda_{\rm G}$ values are available from Ref.~\cite{STAR:2014shf}. Using these inputs, we calculated the corresponding $R_{{\rm free}\;\alpha (o,s,l)}$ values, as shown in Figure~\ref{f:Ro2Rs2Levy}. 

\begin{figure}[t]
	\centering 
	\includegraphics[width=0.70\textwidth]{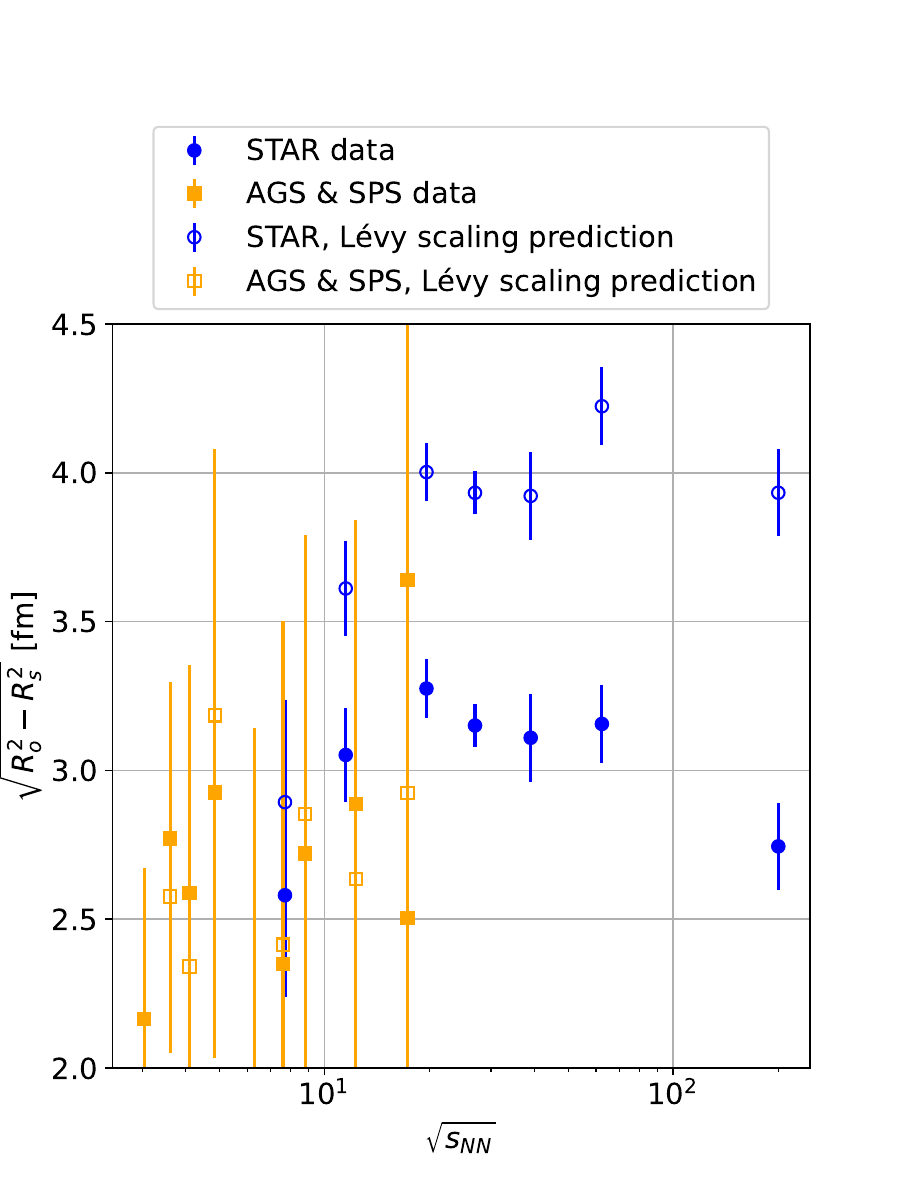}	
	\caption{Scaled Lévy parameter difference $\sqrt{R_{\rm o}^2-R_{\rm s}^2}$, compared to the original Gaussian values from Ref.~\cite{STAR:2014shf}} 
	\label{f:Ro2Rs2diff}%
\end{figure}

As indicated in Refs.~\cite{Chapman:1994yv,Csorgo:2003uv,Lacey:2014wqa}, the combination $\sqrt{R_{\rm o}^2-R_{\rm s}^2}$ for Gaussian sources can be interpreted as a proxy for the emission duration. In the case of Lévy-shaped sources, such identification might not hold anymore. Based on Ref.~\cite{Csorgo:2003uv}, one may consider alternative measures, such as $\left(R_{\rm o}^\alpha-R_{\rm s}^\alpha\right)^{1/\alpha}$. The exact connection between emission duration and femtoscopic radii for Lévy sources is, however, beyond the scope of our manuscript, and we focus on how $\sqrt{R_{\rm o}^2-R_{\rm s}^2}$ depends on collision energy for Lévy-distributed sources. This is shown in Figure~\ref{f:Ro2Rs2diff}. Note that the uncertainty of the estimate based on the validity of the $\widehat{R}$-scaling, being around 4-5\%, is smaller or comparable to the statistical uncertainties of the data.

In line with the expected scaling behavior, the smaller $\alpha$ values observed at larger collision energies increase the radii for the Lévy scaling predictions, and with this, the $out$ and $side$ difference also increases. However, it is interesting to note that the non-monotonic energy dependence is weakened, and the $out-side$ difference appears to be constant above $\sqrt{s_{NN}} = 19$~GeV. These observations demonstrate the need for three-dimensional HBT measurements across collision energies, utilizing Lévy-stable source distributions.

Last but not least, it is important to discuss how 3D HBT radii as parameters of Lévy-distributed sources can be extracted experimentally. One shall start from the correlation function described in Eq.~\ref{e:Cq3D}, containing $R_{\rm o,s,l}$, as well as $\lambda$ and $\alpha$, as parameters. This correlation function can be fitted to experimentally measured 3D correlation functions, similarly to Ref.~\cite{STAR:2014shf} (where a Gaussian source assumption was utilized). Before this can be done, one needs to account for final-state interactions in a self-consistent manner (using the same source for calculating the effect of these interactions as for the correlation function). For 1D measurements, this has been done in Refs.~\cite{PHENIX:2017ino,CMS:2023xyd,NA61SHINE:2023qzr}, using, for example, the calculation package of Ref.~\cite{Nagy:2023zbg}, utilizing spherically symmetric sources. A necessary step in extending this to 3D measurements is to generalize the calculation for sources defined in Eq.~(\ref{e:Levy3D}). Until this becomes available, one may proceed as it was done in Refs.~\cite{STAR:2014shf,Kurgyis:2018zck}, to calculate the correction for final-state interactions under the assumption of an angular averaged, spherically symmetric source. The validity of this approach has been investigated in Ref.~\cite{Kurgyis:2020vbz}. Preliminary results about such measurements with Lévy-stable sources have been reported in Ref.~\cite{Kurgyis:2018zck}, as well as recently by S. Bhosale on behalf of the STAR Collaboration at the 17th Workshop on Particle Correlations and Femtoscopy in 2024.~\footnote{See the mentioned contribution at \href{https://indico.in2p3.fr/event/32030/contributions/144128/}{https://indico.in2p3.fr/event/32030/contributions/144128/}.}

\section{Summary and conclusions}
We demonstrated, based on three example correlation functions, that the variable combination called $\widehat{R}$ is indeed a scaling variable, i.e., it does not strongly depend on the assumed $\alpha$ value. We used this scaling relation to infer HBT radii in the out-side-long system with a free $\alpha$ value from results with a Gaussian assumption ($\alpha=2$). Finally, we investigated the out-side difference as a proxy for emission duration. Our results indicate that the trend of this difference is strongly affected by the assumption for the shape, and unlike the case of a Gaussian assumption, there may be no non-monotonic behavior of the out-side difference as a function of collision energy. This underlines the importance of performing three-dimensional experimental HBT measurements with an unrestricted $\alpha$ parameter.

\section*{Acknowledgements}
This research was funded by the NKFIH grants TKP2021-NKTA-64, PD-146589, K-138136 and K-146913.

\begin{appendices}
\section{Appendix A}
\begin{figure}
	\centering 
	\includegraphics[width=0.95\textwidth]{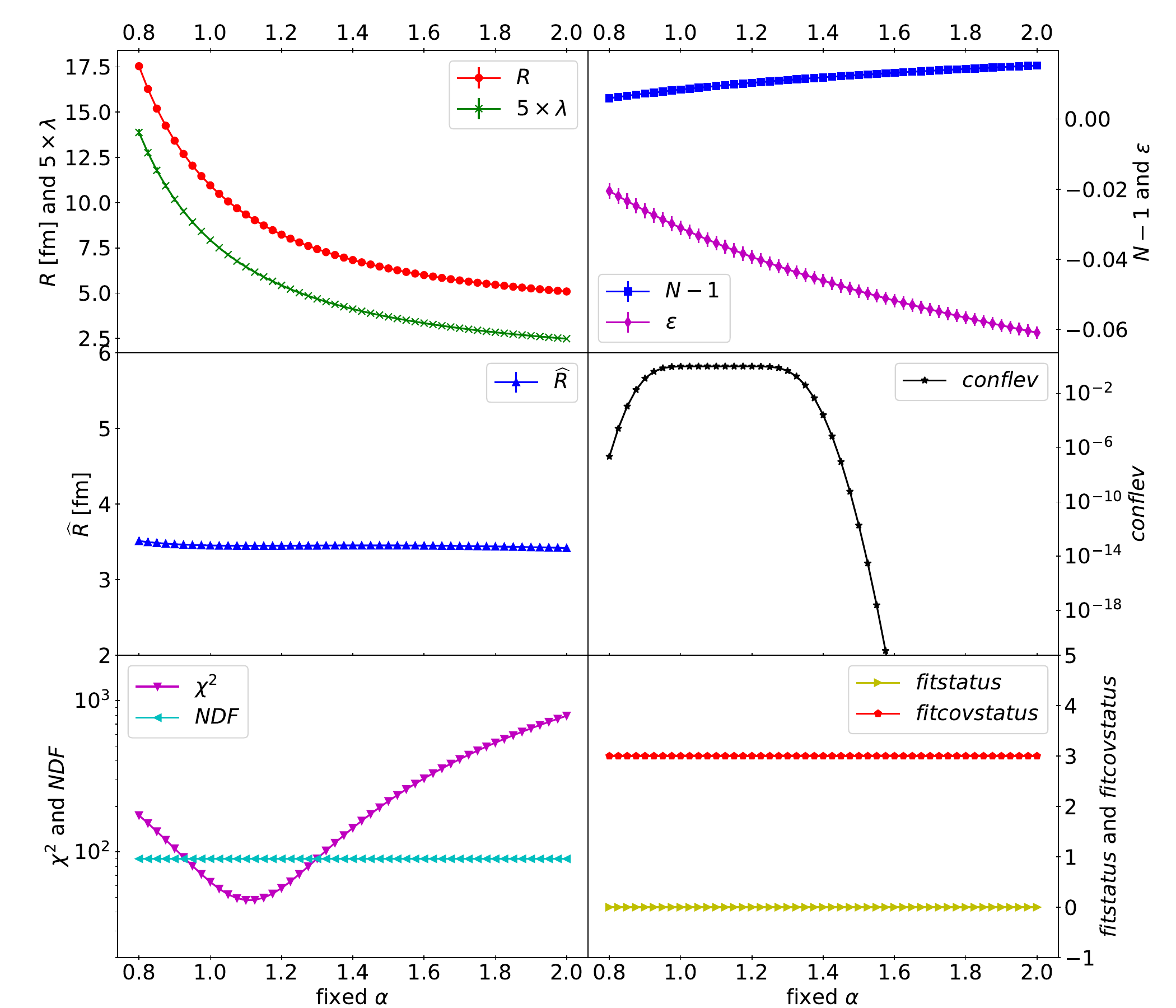}	
	\caption{Fit parameters and quality from various fixed-$\alpha$ fits to correlation function data of Figure 3a from Ref.~\cite{PHENIX:2017ino}. The plots show parameters $R$ and $\lambda$, fit normalization parameter $N$ and residual slope $\epsilon$, scaling parameter $\widehat{R}$, confidence level, $\chi^2$ and NDF, as well as the ``fit status'' and ``covariance matrix status'' parameters.}
	\label{f:HEPData-ins1624209-v1-Figure_3a}%
\end{figure}

\begin{figure}
	\centering 
	\includegraphics[width=0.95\textwidth]{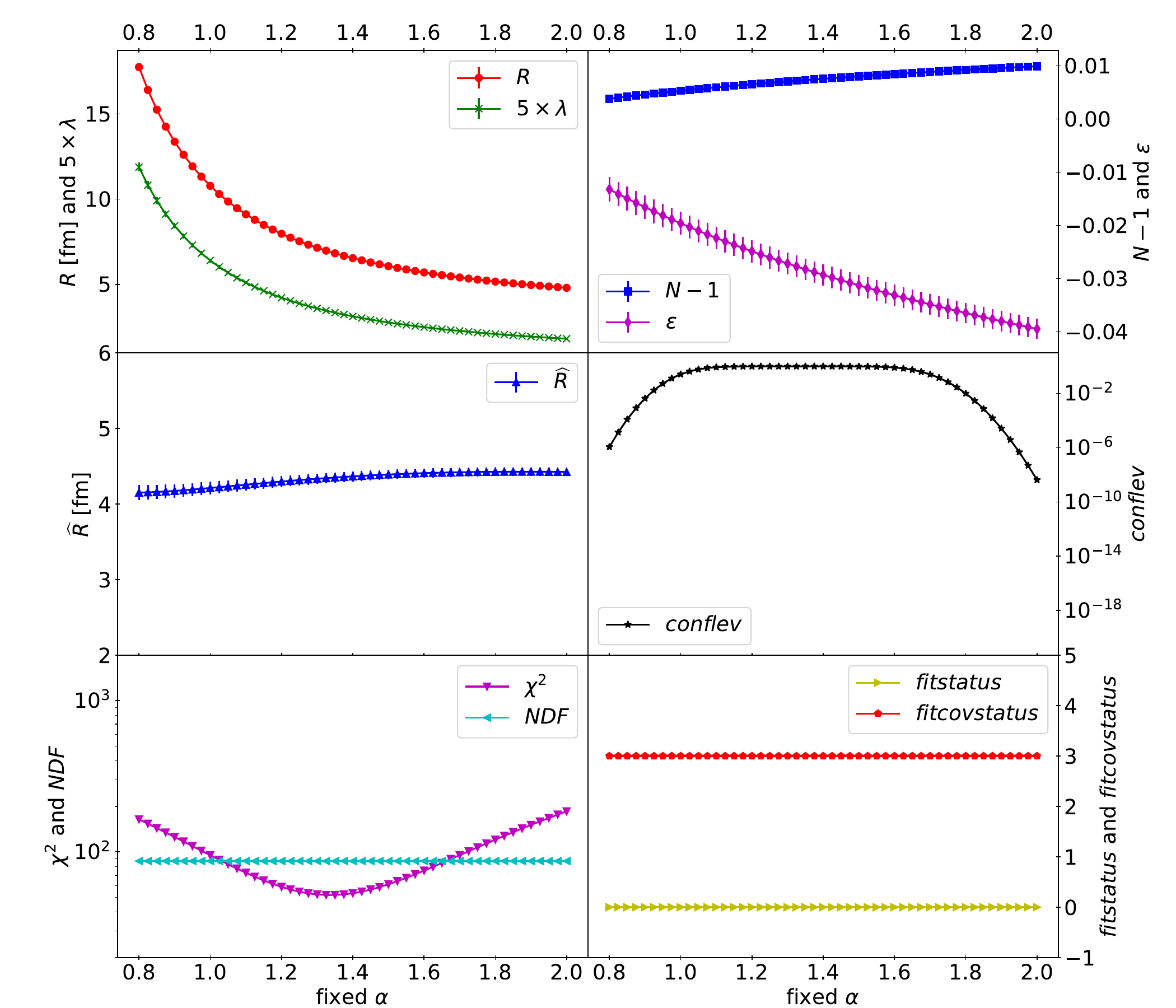}	
	\caption{Fit parameters and quality from various fixed-$\alpha$ fits to correlation function data of Figure 3b from Ref.~\cite{PHENIX:2017ino}, similarly to Fig.~\ref{f:HEPData-ins1624209-v1-Figure_3a}.} 
	\label{f:HEPData-ins1624209-v1-Figure_3b}%
\end{figure}

\begin{figure}
	\centering 
	\includegraphics[width=0.95\textwidth]{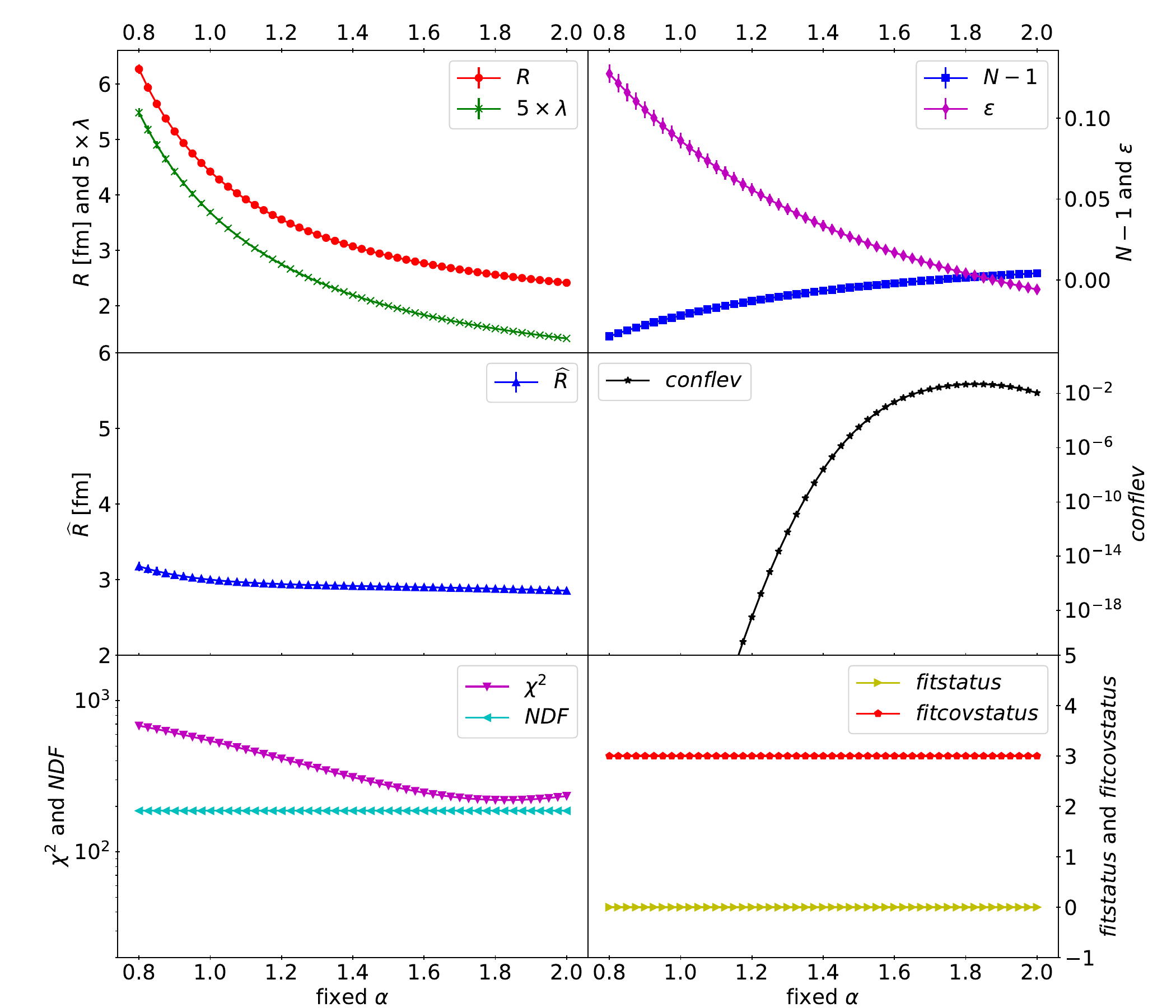}	
	\caption{Fit parameters and quality from various fixed-$\alpha$ fits to correlation function data of Figure 2 from Ref.~\cite{CMS:2023xyd}, similarly to Fig.~\ref{f:HEPData-ins1624209-v1-Figure_3a}.} 
	\label{f:HEPData-ins2670243-v1-Table_2}%
\end{figure}

In this appendix we show fits with a range of possible fixed $\alpha$ values to data obtained in $\sqrt{s_{NN}}=200$ GeV Au+Au collisions.~\cite{PHENIX:2017ino} Figs. \ref{f:HEPData-ins1624209-v1-Figure_3a}, \ref{f:HEPData-ins1624209-v1-Figure_3b}, and \ref{f:HEPData-ins2670243-v1-Table_2} show all fit parameters ($R$, $\lambda$, $N$, $\epsilon$) and fit quality indicators ($\chi^2$, NDF, confidence level, fit status and covariance matrix status), along with the scaling variable $\widehat{R}$, for fixed $\alpha$ values from 0.8 to 2.0. Here NDF means the number of degrees of freedom (number of data points minus the number of free fit parameters), confidence level means the p-value of the fit (calculated from $\chi^2$ and NDF), fit status and covariance matrix status come from the Minuit optimization library~\cite{James:1975dr}, and their best values are 0 and 3, respectively (indicating a fully converged fit and fully positive definite covariance matrix). These fit quality indicators are important as they demonstrate the good quality of the fits. Without them, one cannot be sure that the best fit parameters indeed do represent the data. These plots show that (especially for larger $\alpha$, above $\sim$1.4) the scaling variable $\widehat{R}$ is indeed approximately independent of the assumed fix $\alpha$ value, hence can be used to infer $R$ at a given $\alpha$ from $R$ at a different $\alpha$. Note also that clearly, $\alpha$ values close to 2.0 provide a statistically inferior description of the data, giving another reason for performing fits with a released $\alpha$.

\end{appendices}

\bibliography{main}

\end{document}